\documentclass[11pt]{article}
\usepackage{amsmath}
\usepackage{amsfonts}
\usepackage{amssymb}
\usepackage{amsthm}
\usepackage{epsfig}
\usepackage{graphicx}
\usepackage[ruled,vlined]{algorithm2e}
\usepackage{hyperref}
\usepackage[all]{hypcap}
\usepackage{subfig}
\usepackage{graphicx}
\usepackage[countmax]{subfloat}
\usepackage{setspace}
\usepackage{subfig}
\usepackage{gensymb}
\usepackage{color}
\usepackage{bm}
\usepackage{authblk}
\usepackage{fullpage}
\usepackage{url}
\usepackage[authoryear]{natbib}
\usepackage{etoolbox}
\usepackage{amsthm}
\usepackage{fancyhdr}
\makeatletter

\newif\ifabbreviation
\pretocmd{\thebibliography}{\abbreviationfalse}{}{}
\AtBeginDocument{\abbreviationtrue}

\usepackage{lastpage}
\begin{document}

%\pagestyle{empty}
%\begin{center}

\title{A Spatio-Temporal Modeling Approach for Weather Radar Reflectivity Data and Its Applications in Tropical Southeast Asia}
%\large (submitted to the Applications and Cast Studies Track \\ Journal of the American Statistical Association) }
%\author[1]{}
\author[1]{Xiao Liu} %\thanks{A.A@university.edu}}
\author[2]{Viknesswaran Gopal} %\thanks{B.B@university.edu}}
\author[1]{Jayant Kalagnanam}
\affil{IBM T.J. Watson Research Center}
\affil[2]{Department of Statistics and Applied Probability, National University of Singapore}
\renewcommand\Authands{ and }
\date{ }

\maketitle

\vspace{0.5cm}
%
%Research Team \footnote{\noindent IBM Research}
%and
%

%\today

%\end{center}

\begin{abstract}
Weather radar echoes, correlated in both space and time, are the most important input data for short-term precipitation forecast. Motivated by real datasets, this paper is concerned with the spatio-temporal modeling of two-dimensional radar reflectivity fields from a sequence of radar images. Under a Lagrangian integration scheme, we model the radar reflectivity data by a spatio-temporal conditional autoregressive process which is driven by two hidden sub-processes. The first sub-process is the dynamic velocity field which determines the motion of the weather system, while the second sub-process governs the growth or decay of the strength of radar reflectivity. The proposed method is demonstrated, and compared with existing methods, using the real radar data collected from the tropical southeast Asia. Note that, since the tropical storms are known to be highly chaotic and extremely difficult to be predicted, we only focus on the modeling of reflectivity data within a short-period of time and consider the short-term prediction problem based on the proposed model. This is often referred to as the nowcasting issue in the meteorology society.
\end{abstract}

\noindent\textbf{Key words:} {\em Spatio-Temporal Statistics, Radar Image, Space-Time Conditional Autoregressive, Precipitation Forecast}

\clearpage

\section{Introduction} \label{sec:Intro}

Rain and snow return echoes on radar images. Since the discovery of such a phenomenon during the World War II, many methods have been developed to locate, track and predict precipitation, with the fundamental idea in common being the \emph{spatio-temporal extrapolation/advection} of radar reflectivity (echo) field. Nowadays, these methods are collectively known as the radar-based Quantitative Precipitation Forecasts (QPF) in the meteorological society (Wilson \emph{et al.} 1998, RMI 2008). From the observed radar reflectivity, precipitation intensity can be estimated through reflectivity-rainfall relationships (Testik and Gebremichael 2013); for example, the well-known Marshall-Palmer relationship between radar reflectivity and the distribution of the size of rain drops (Marshall and Palmer 1948). Statistical spatio-temporal approaches have also been proposed for the calibration of radar-rainfall data. For example, Brown \emph{et al.} (2001) developed a high dimensional multivariate state space time series model for the space-time calibration of radar-rainfall data. Fuentes \emph{et al.} (2008) introduced a framework, based on the spatial logistic regression model, to combine radar reflectivity and gage rainfall data, by expressing the different sources of rainfall information in terms of an underlying unobservable spatial temporal process with the true rainfall values. Xu \emph{et al.} (2005) proposed a kernel-based radar reflectivity nowcasting approach that efficiently parameterizes spatio-temporal dynamic models in terms of integro-difference equations within a hierarchical framework. The kernel-based method has its advantages in incorporating the underlying physical knowledge about the rainfall process. A good example is presented by Sigrist \emph{et al.} (2012) in which the authors presented a hierarchical Bayesian model for short-term predictions of rainfall, which is based on a temporal autoregressive convolution with spatially colored and temporally white innovations.
Compared with pure physical relationships, statistical models are often more effective in capturing the uncertainty and dynamics of the radar-rainfall relationship.

This paper is concerned with the spatio-temporal modeling of the radar reflectivity data, i.e., a sequence of radar images, collected from the tropical southeast Asia. In particular, we focus on the modeling of reflectivity data within a short-period of time, say, 10 to 30 minutes, and consider the short-term prediction problem based on the proposed model. This is often referred to as the nowcasting issue in the meteorology society.
Since the tropical storms are known to be highly chaotic, additional input and methodologies are required for the long-term modeling and prediction, which is out of the scope of the paper. Two fundamental issues are to be addressed in this paper: the construction of the velocity vector field associated with the weather system, and the modeling of the growth and delay of the reflectivity level.

The radar reflectivity field of a weather system (for example, a hurricane or tropical storm) is usually moving on the radar screen. The motion or velocity vector field, however, is not directly observable from radar images. In the meteorological literature, both the gradient-based (i.e., intensity-invariant) and pattern-based (i.e., shape-invariant) methods have been used for estimating the velocity field. The gradient-based methods, proposed by Horn and Schunck (1981) in studying the two-dimensional velocities of the brightness patterns of image sequences, assume that the reflectivity is invariant over short time intervals, and obtain the velocity vectors based on the first-order variation of the reflectivity field. Although often known as the Optical Flow in computer science, the gradient-based method is referred to as the method of Lagrangian Persistence by meteorologists, and is now the building block of many storm-tracking algorithms (Bowler \emph{et al.} 2004). The pattern-based method, on the other hand, is based on the concept of area tracking. Tracking areas (i.e., radar image pixel arrays) are defined around all pixel grid points, and corresponding areas are searched in the next radar image by maximizing the cross-correlation between areas. Then, the velocity field can be constructed given the spatial lags between areas and the time lag between two radar scans. Unlike the gradient-based methods, the pattern-based methods assume that the shape of the reflectivity patterns within defined areas do not change over short time intervals. In the literature, the pattern-based method was firstly introduced by Leese \emph{et al.} (1971) to identify cloud motion from satellite images, and has later been successfully used for the nowcasting of precipitation with radar over a complex orography (Rinehart and Garvey 1978, Li \emph{et al.} 1995, Li and Lai 2004). The construction of the velocity vector field serves as a \emph{preliminary} step for the spatio-temporal modeling approach to be described in this paper, and readers may refer to Gelpke and Kunsch (2001) for a comprehensive review of the statistical methods for motion estimation. In Appendix, we also provide necessary preliminaries on the construction of the velocity field.

The strength of radar reflectivity can grow or decay during the collision-coalescence process due to vertical and horizontal winds, breakup and evaporation. In tropical areas considered in this paper, the growth and decay of the reflectivity becomes more prominent due to the presence of many small-scale convective storm cells embedded in the storm system. These convective cells usually change rapidly and their lifetime can be as short as tens of minutes, posing a tremendous change to the modeling of the radar reflectivity data of tropical storms. In 1963, the American meteorologist Edward Lorenz had already pinned down the chaotic nature of atmospheric convection for weather-forecasting. It has been found that errors in the linear extrapolation of the radar echo field assuming persistent reflectivity level are mainly due to the growth and decay of the reflectivity (Browning \emph{et al.} 1982). The prediction of the growth and decay, however, is a not trivial task and most of the early efforts have not been very successful due to the nature of the problem (Wolfson \emph{et al.} 1999). In fact, most of the existing operational QPF systems in the world do not takes into account the growth and decay phenomenon, including the GANDOLF system developed in UK (Bowler \emph{et al.} 2004) and the SWIRLS system developed in Hong Kong (Li and Lai 2004). One of the operational systems that does consider the growth and decay of the reflectivity is the McGill Algorithm for Precipitation Nowcasting by Lagrangian Extrapolation (MAPLE) (Radhakrishna \emph{et al.} 2012, Germann and Zawadzki 2002). It is pointed out by the authors that the growth and decay of precipitating systems can be estimated by the mismatch between two radar scans under the Lagrangian integration scheme. However, the authors assume that the growth and decay are persistent over time, and a statistical model that incorporates both the spatial and temporal correlation of the reflectivity data is not available.
%Note that, the modeling of the growth and decay of reflectivity for short-term precipitation forecast is not always necessary when the weather system is less dynamic, and it seems sufficient to track the movement of the reflectivity field (add existing systems).

This paper proposes a spatial-temporal modeling approach for weather radar reflectivity fields under a Lagrangian integration scheme. In particular, we consider a classic forced-advection problem under the Lagrangian frame of reference, where an observer watches the world evolves around her as if she traveled with the radar image pixel arrays within a velocity field. Based on the discrete approximation to the forced-advection problem, the growth and decay of the reflectivity is defined as the mismatch between two radar scans given the constructed velocity vector field. We model the reflectivity growth and decay by extending the Spatial Temporal Conditional Auto-Regressive (STCAR) model (Mariella and Tarantino 2010). The Conditional Regressive Model (CAR) model, or Gauss-Markov model, has been extensively used for modeling geographical area data when a spatial phenomenon at a location is affected by its neighboring areas (Cressie 1993, Besag and Kooperberg  1995, Stern and Cressie 2000, Carlin and Banerjee 2003, Banerjee \emph{et al.} 2004). The spatial structure implied by the CAR model has been investigated by Wall (2004). In our context, a radar image pixel array (see Section \ref{sec:Data}), which consists of a number of image pixels, can be naturally modeled as a spatial area. And the reflectivity growth on any pixel array is almost always affected by the growth of the reflectivity on its neighboring pixels. Because the reflectivity is recorded at each spatial location in a time interval, the STCAR model, which is essentially an autoregressive model for a temporal sequence of CAR, takes into account both the spatial dependence of the reflectivity growth among neighboring image pixel arrays and the temporal dependence of the reflectivity growth on the same pixel array. Unlike many other conventional multivariate geographical area data models, the locations of pixel arrays keeps changing over time due to the motion of the reflectivity field. Hence, the distance and the spatial relationship between two pixels arrays are no longer time-invariant, making the structure of spatial dependence more dynamic and complicated. On the other hand, it is also desirable to keep the model simple and computationally efficient. Note that an STCAR model can be directly specified by choosing the joint distribution of a sequence of Markov random fields via conditional and marginal distributions. Based on the recent findings of Radhakrishna \emph{et al.} (2012), the growth and decay of the precipitation intensity over short time intervals are usually well approximated by Gaussian. This finding, which is also verified in our application example presented in Section \ref{sec:case}, allows us to conveniently specify the conditional and marginal distributions for the STCAR model. Furthermore, compared to other multivariate areal data model, such as the Generalized Multivariate Conditional Auto-Regressive (GMCAR) model (Jin \emph{et al.} 2005), the STCAR model provides a simple yet efficient way to capture the spatial association at a particular time point by a single parameter, leading to a significant reduction in the computational time. Note that, as a new weather radar image typically becomes available every 5 minutes, the model needs to be constructed or updated in a few minutes considering the time consumed by data processing and transfer.

The paper is organized as follows. Section \ref{sec:Data} describes the radar data and the basic settings of a radar image. The general modeling framework and details are provided in Section \ref{sec:STCAR}. In Section \ref{sec:case}, an application example are presented using real radar data. Section \ref{sec:conclusion} is the conclusion of the paper.

\section{Motivating Example, Data, and Basic Settings} \label{sec:Data}
In this section, we provide an introduction to the motivating example and the weather radar data used in this paper. In particular, the basic settings of a weather radar image is described.

\subsection{Singapore Floods}
As a tropical island country located 137km north of the equator, Singapore has a tropical rainforest climate with abundant rainfall. Based on a 145-year survey from the year 1869 to 2013, the average yearly rainfall of Singapore is approximately 2344mm, and the average raining days of a year is 178 days (NEA 2014). The yearly rainfall of Singapore is respectively 4 and 2.5 times higher than that of London and Seattle.

The 2010-2013 Singapore floods refers to the series of flash floods that hit various parts of the city state Singapore since 2010. The floods came about due to the higher-than-average rainfall that aggregated over a short period of time. Detailed descriptions and a complete list of flooding incidents are available from Wikipedia under ``2010-2013 Singapore floods''.

\begin{figure}
\begin{center}
 \includegraphics[width=1\textwidth]{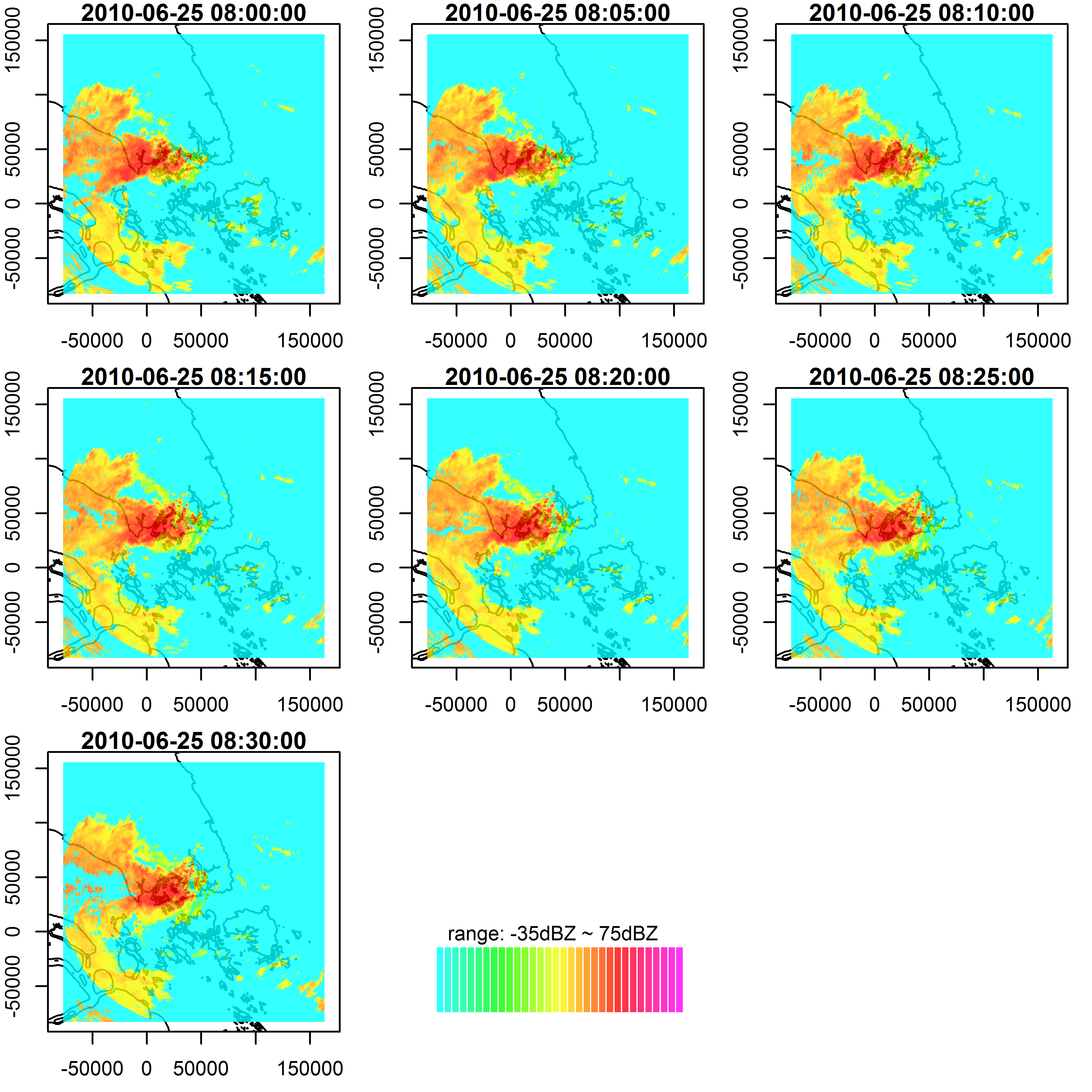} % regenerate
\caption{Radar scans of the tropical storm on the morning of 25 June 2010} \label{fig:dBZcase}
\end{center}
\end{figure}

In this paper, we focus on the heavy rain event on the early morning of 25 June 2010. A torrential downpour early that morning triggered flash floods across Singapore. The flood also caused morning rush hour traffic to come to a virtual standstill on all major expressways. It was reported that the heavy downpour on that Friday morning was equivalent to about 60 percent of the average monthly rainfall in June. Social media and micro-blogging sites such as Facebook and Twitter were awash with flood photos and users exchanging pictures. Figure \ref{fig:dBZcase} shows 7 consecutive weather radar scans of reflectivity taken from 8AM to 8:30AM, 25 June 2010. A tropical storm, moving from the west to the east, is clearly identified by the high-reflectivity area (the orange area) in the figures. Note that, Singapore is the small island at the center of the image, and the x- and y-coordinates are based on the SVY21 projection which has a one-to-one correspondence with the latitude-longitude projection system. %From the precipitation forecast point of view, we only show positive reflectivity values (in units of dBZ) in the figure because negative reflectivity values do not lead to any precipitation and thus are not of our interest.
\begin{figure}
\begin{center}
%\subfloat[]{
%\includegraphics[width=0.6\textwidth]{figures/ScanExample.pdf}}\\
%\subfloat[]{
\includegraphics[width=0.6\textwidth]{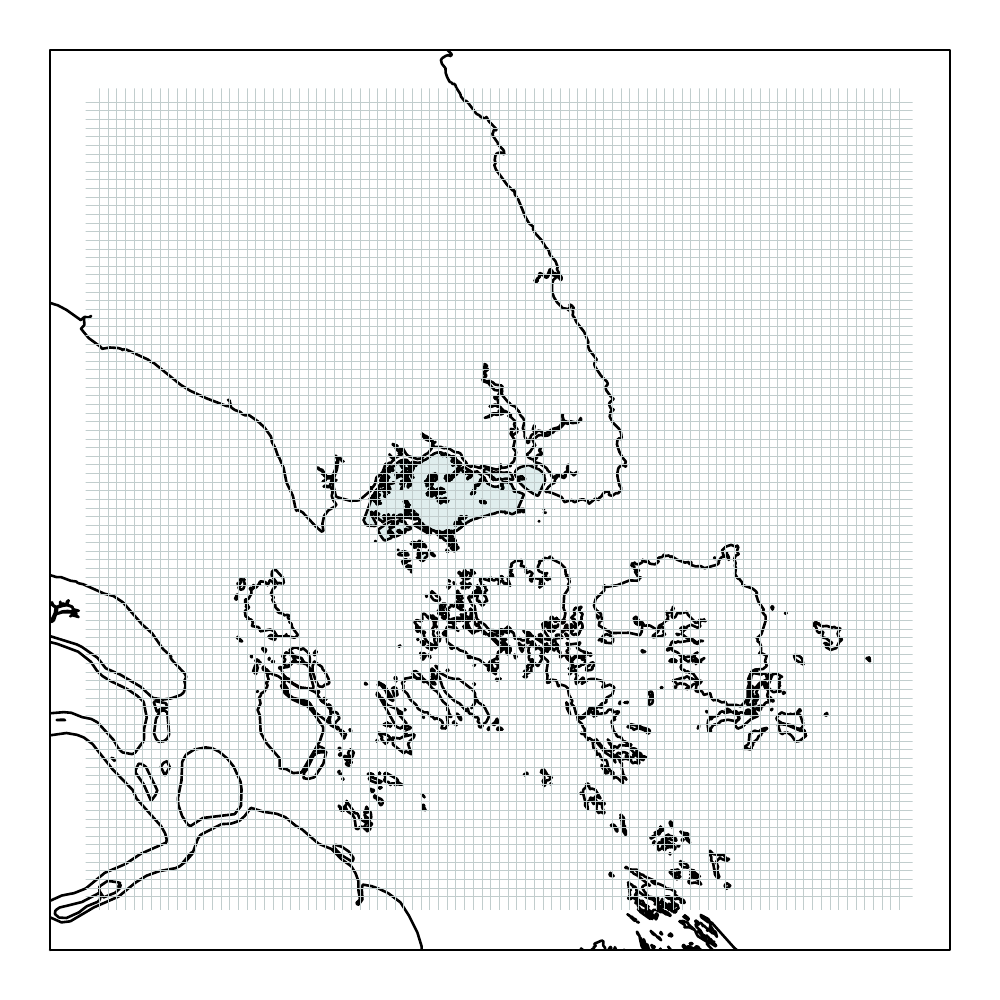}
%\caption{(a) A CAPPI reflectivity image taken at 1km above the mean sea level and at a particular time (in units of dBZ); (b) Centers of the 8649 pixels arrays that occupy the spatial domain of a CAPPI image} \label{fig:2}
\caption{Centers of the 8649 pixels arrays that occupy the spatial domain of a CAPPI reflectivity image} \label{fig:pixelarrays}
\end{center}
\end{figure}

\subsection{Weather Radar Data and Basic Settings}
The radar images used in the paper are generated by a dual polarization Meteorological Doppler Weather Radar (MDWR) located at the Singapore Changi Airport--the eastern tip of the main island.
%Figure \ref{fig:1} displays the map of Singapore as well as the location of the weather radar system.
%\begin{figure}
%\begin{center}
%%\subfloat[]{
%%\includegraphics[width=1\textwidth]{figures/Changi.pdf}}
%%\subfloat[]{
%%\includegraphics[width=0.276\textwidth]{figures/Radar.png}}
%    \includegraphics[width=0.5\textwidth]{figures/Changi.pdf}
%    \caption{Map of Singapore and the location (indicated by the red dot) of the meteorological doppler weather radar} \label{fig:1}
%\end{center}
%\end{figure}
A modern MDWR system generates hundreds of products related to meteorological conditions. In this paper, we use the standard Constant Altitude Plan Position Indicator (CAPPI) \emph{reflectivity} data at 1km above the mean sea level. The method can be applied to other CAPPI datasets at other user-definable heights. CAPPI reflectivity images contain information about the reflectivity (in units of dBZ) at given grid points and times. These images are taken at 5min interval and each image is arranged on a Cartesian 2D grid of $480\times480$ pixels, with the top-left and lower-right corners given by $(102.892^\circ \text{E}, 2.42799^\circ \text{N})$ and $(105.052^\circ \text{E}, 0.269748^\circ \text{N})$, respectively. The resolution of each grid square is approximately $0.5\times0.5$ kilometers. %Figure \ref{fig:dBZcase} shows the reflectivity images at 1km above the mean sea level.

The $480\times480$ pixels of a radar scan are further divided into $93\times93$ pixel arrays (i.e., boxes or tiles). Each pixel array has a fixed size of $19\times19$ pixels and covers an area of $90.25\mathrm{km}^2$. The centers of the pixel arrays are spaced 5 pixels apart, which is approximately $2.5\mathrm{km}$. Note that, neighboring pixel arrays overlap with each other. Figure \ref{fig:pixelarrays} plots the centers of the $93\times93$ pixel arrays that occupy the entire spatial domain of a radar scan. This particular setting of pixel arrays is commonly adopted by the meteorological society (Li and Lai 2004).

\section{A Spatio-Temporal Model} \label{sec:STCAR}
\subsection{The Framework} \label{sec:model}
Consider a Lagrangian-type of advection scheme within which an observer watches the world evolves around her as if she traveled with the pixel arrays. Let $\mathbb{S} \subseteq \mathbb{R}^2$ and $\mathbb{T} \subseteq \mathbb{N}^+$ be the spatial and temporal domains, we denote by $x_{i,t}$ the location of pixel array $i$ at time $t$, $i=1,2,...,n$. Note that, pixel arrays are advected by a velocity (wind) field as the weather system moves. In addition, we also denote by $s_i \in \mathbb{S}$, $i=1,2,...,n$, the initial location of pixel array $i$ at some reference time $t=1$. Hence, $x_{i,t}=s_i$ for all $i$ when $t=1$.

%Since each pixel array contains $19\times19$ pixels, we let a random function $Y_t(k_1,k_2;s_i)$ represent the reflectivity at time $t$ on the $(k_1,k_2)\text{th}$ pixel within the $i\text{th}$ pixel array centered at $s_i$, where $k_1,k_2=1,2,...,19$ and $t \in \mathbb{N}^+$.
The velocity field characterizes the motion of pixel arrays. Since a two-dimension problem is considered in this paper, the velocity field is the projection of the actual wind field on the surface 1km above the sea level. Let $\textbf{x}_t=(x_{1,t},x_{2,t},...,x_{n,t})^T$ be a collection of the locations of all pixel arrays at time $t$, we define the forward translation (i.e., shift) operator $\Gamma$ as:
\begin{eqnarray}
   \Gamma\textbf{x}_t = (\Gamma x_{1,t},\Gamma x_{2,t},...,\Gamma x_{n,t})^T= \textbf{x}_{t+1}
   \label{eq:1}
\end{eqnarray}
The translation operator $\Gamma$ can be fully determined given the velocity field estimated from a series of consecutive radar scans, and we leave the details to Appendix \ref{sec:appendix}. Based on equation (\ref{eq:1}), the $m\text{-step}$ forward translation operator and inverse translation operator are also defined:
\begin{eqnarray}
   \underbrace{\Gamma\Gamma...\Gamma}_{m}=\Gamma^{(m)}{x}_t = \textbf{x}_{t+m}, \quad m \in \mathbb{N}^+
   \label{eq:2}
\end{eqnarray}
\begin{eqnarray}
   \Gamma^{(-m)}\textbf{x}_{t} = \textbf{x}_{t-m}, \quad m \in \mathbb{N}^+.
   \label{eq:3}
\end{eqnarray}

Similarly, let $\textbf{Z}_t=\textbf{Z}(\textbf{x}_t)=(Z(x_{1,t}),Z(x_{2,t}),...,Z(x_{n,t}))^T$ denote the radar reflectivity on all pixel arrays at time $t$ with $Z(x_{i,t})$ being the reflectivity on pixel array $i$ ($i=1,2,...,n$). Then, a classic forced-advection model is defined as (Staniforth and Cote 1991):
\begin{eqnarray}
   \frac{d\textbf{Z}_t}{dt} = \frac{\partial\textbf{Z}_t}{\partial t} + \textbf{v}_t \cdot \nabla \textbf{Z}_t = \textbf{G}_t
   \label{eq:4}
\end{eqnarray}
where $\textbf{v}_t$ is the velocity field at time $t$ over the space, and $\textbf{G}_t=\textbf{G}(\textbf{x}_t)=(G(x_{1,t}),G(x_{2,t}),...,G(x_{n,t}))^T$ is the forcing term that represents the growth (or, decay) of the radar reflectivity on pixel arrays at time $t$. Here, $G(x_{i,t})$ is the reflectivity growth on pixel array $i$.

It is well-known that there exists a discrete approximation to the forced-advection problem (\ref{eq:4}) as follows:
\begin{eqnarray}
\begin{split}
   \textbf{Z}_{t+1} = \textbf{Z}(\textbf{x}_{t+1}) & = \textbf{Z}(\Gamma^{(-2)}\textbf{x}_{t+1}) + 2\textbf{G}(\Gamma^{(-1)}\textbf{x}_{t+1}) + O(\Delta^2) \\ & \approx \textbf{Z}_{t-1} + 2\textbf{G}_{t}
   \label{eq:approximation}
\end{split}
\end{eqnarray}
where $\Delta$ is the length of time intervals.

The advection scheme (\ref{eq:approximation}) serves as the basis of our model, which suggests that the process $\{\bm{Z}_t\}$ is driven by \emph{two hidden sub-processes}. The first sub-processes is the dynamic velocity field which determines translation operator $\Gamma$, while the second is the reflectivity growth process $\{\bm{G}_t\}$. In other words, given the trajectory of any pixel array $i$, the reflectivity on that pixel array at time $t+1$ is the sum of the reflectivity on the pixel array at time $t-1$ and two times the reflectivity growth on the same pixel array at time $t$. Note that, neither the velocity field (i.e., the trajectories of pixel arrays) nor the reflectivity growth are directly observed.

To track the pixel arrays at different times, $(\textbf{x}_1,\textbf{x}_2,...,\textbf{x}_t)$,
%and obtain the reflectivity $(\textbf{Z}_1,\textbf{Z}_2,...,\textbf{Z}_t)$ associated with the tracked arrays,
the velocity field needs to be constructed. The velocity vector field of the weather system can be obtained by analyzing two consecutive radar scans. Once the velocity field has been established, the translation operators are fully defined and the locations of pixel arrays at different times can be tracked. Hence, the construction of the velocity field serves as a preliminary step for the entire modeling approach. As discussed in Section \ref{sec:Intro}, we adopt one popular pattern-based method known by the meteorological society as the TREC method (Tracking Radar Echoes by Correlation). Since the TREC method has been well explained by Rinehart and Garvey (1978), Li \emph{et al.} (1995), and Li and Lai (2004), we provide a brief review of the principle of the method in Appendix \ref{sec:appendix}. In this paper, we treat the translation operators as known from the constructed velocity vector field. The error associated with the estimated velocity field is absorbed by the statistical model of $\{\bm{G}_t\}$ to be described.

Once velocity field has been constructed, it becomes possible to track the locations of pixel arrays at different times, $(\textbf{x}_1,\textbf{x}_2,...,\textbf{x}_t)$, and obtain the observed reflectivity $(\textbf{Z}_1,\textbf{Z}_2,...,\textbf{Z}_t)$ associated with the tracked pixel arrays from radar images. Then, the reflectivity growth process $\{\bm{G}_t\}$ is obtained from the Lagrangian-type of advection scheme (\ref{eq:approximation}):
\begin{eqnarray}
\begin{split}
    \textbf{G}_{t} = \textbf{G}(\Gamma^{(-1)}\textbf{x}_{t+1})  = \frac{\textbf{Z}(\textbf{x}_{t+1}) - \textbf{Z}(\Gamma^{(-2)}\textbf{x}_{t+1})}{2}
   \label{eq:growth}
\end{split}
\end{eqnarray}
In other worlds, the reflectivity growth is exactly the mismatch between two radar scans under the velocity vector field. In a special case in which pixel arrays do not move, we would be able to obtain the reflectivity growth immediately from the mis-match or difference of the reflectivity from two consecutive radar scans. Of course, this is never the case in reality when modeling the radar reflectivity data.

In the next sub-section, we describe how the reflectivity growth process $\{\bm{G}_t\}$ is modeled.
%we consider a statistical model that describes the process $\{ \textbf{Z}_{t}; t \in \mathbb{T} \}$ as an auto-regressive process with shift:

%\begin{eqnarray}
%\begin{split}
%   \textbf{Z}_{t+1} &  = \textbf{Z}(\Gamma^{(-2)}\textbf{x}_{t+1}) + 2\textbf{G}(\Gamma^{(-1)}\textbf{x}_{t+1}) \\ & \approx \textbf{Z}_{t-1} + 2 (\mu_{t} + \textbf{Y}_{t} )
%   \label{eq:statFrame1}
%\end{split}
%\end{eqnarray}
%where $\mu_{t}$ captures the deterministic large-scale spatial trend of the reflectivity growth, and $\textbf{Y}_{t}$ describes the small-scale variation of the reflectivity growth with zero-mean and spatial covarance $\mathbf{\Sigma}_{{\bm{Y}_t}}$. In Section \ref{sec:stcar}, we provide details of how $\mu_{t}$ and the STCAR model for $\textbf{Y}_{t}$ are constructed.

%Suppose that $\textbf{S} \subseteq \mathbb{R}^2$ and $T \in \mathbb{N}^+$ respectively be the spatial and temporal domains.

\subsection{The Modeling} \label{sec:model2}

Extrapolating radar echoes is the mainstay of nowcasting, i.e., the prediction of precipitation in a very near future. In Singapore, for example, 30-minutes-ahead rainfall prediction is often of interest and can be made from analyzing radar images. Note that, given the extremely dynamic nature of tropical storms, additional information is often needed (e.g., satellite and Numerical Weather Prediction (NWP) model output) in order to extend the time period of nowcasting. Since only radar images are used in this paper, it is a natural choice to consider the (possibly transformed) reflectivity growth process $\{\bm{G}_t\}$ as an Autoregressive (AR) process. In particular, we assume that the process $\{\bm{G}_t\}$ has the following general form:
\begin{eqnarray}
     \bm{B}_{t}\textbf{G}_{t} = r_1 \bm{B}_{t-1}\textbf{G}_{t-1} + r_2 \bm{B}_{t-2}\textbf{G}_{t-2} + ... + r_q \bm{B}_{t-q}\textbf{G}_{t-q} + \epsilon_t
   \label{eq:AR}
\end{eqnarray}
where $(r_1,r_2,...,r_q)$ are the autoregression coefficients, $\epsilon_t$ is some spatio-temporal error process, $q$ is the order of the AR process, and the matrix $\bm{B}$ is to be determined for each time point.

At any time $t$, we model $\textbf{G}_{t}$ as a spatio-temporal process and assume that $\textbf{G}_{t}$ can be expressed by:
\begin{eqnarray}
     \textbf{G}_{t} = \mu_t + \textbf{Y}_{t}
   \label{eq:Gt}
\end{eqnarray}
where $\mu_t = (\mu_{1,t}, \mu_{2,t}, ..., \mu_{n,t})^T$ captures the deterministic large-scale spatial trend of the reflectivity growth at time $t$ with its element $\mu_{i,t}$ representing the mean reflectivity growth of the pixel array $i$, and $\textbf{Y}_{t}$ describes the small-scale random variation of the reflectivity growth with zero-mean and spatial covarance $\mathbf{\Sigma}_{t}$.

In particular, the mean function $\mu_t$ needs to be flexible enough to handle the complex reflectivity growth over the spatial domain. Hence, we consider a locally weighted mixture of linear regression model as below (Stroud \emph{et al.} 2001):
\begin{eqnarray}
     \mu_{i,t}  =  \sum_{j=1}^{J}\pi_{j}(x_{i,t})\textbf{f}_j^T(x_{i,t})\gamma_{j,t}
   \label{eq:MeanFunction}
\end{eqnarray}
where $\textbf{f}_j(x_{i,t})$ is a column vector of known basis functions, $\gamma_{j,t}$ are column vectors of unknown parameters at time $t$, and $\pi_j$ is a non-negative kernel centered at chosen locations. Let $\textbf{F}_{j,t}=(\textbf{f}_j(x_{1,t}), \textbf{f}_j(x_{2,t}), ..., \textbf{f}_j(x_{n,t}))$ and $\pi_j = (\pi_{j}(x_{1,t}), \pi_{j}(x_{2,t}), ..., \pi_{j}(x_{n,t}))$, we have
\begin{eqnarray}
     \mu_{t}  =  \textbf{F}_t\gamma_{t},
   \label{eq:MeanFunction}
\end{eqnarray}
where $\textbf{F}_t=( \textrm{diag}(\pi_1)\textbf{F}_{1,t}, \textrm{diag}(\pi_2)\textbf{F}_{2,t}, ..., \textrm{diag}(\pi_J)\textbf{F}_{J,t})$, and $\gamma_{t}=(\gamma_{1,t}^T,\gamma_{2,t}^T,...,\gamma_{J,t}^T)^T$.

The second term, $\textbf{Y}_{t}$, in (\ref{eq:Gt}) is modeled by an STCAR model to handle both the spatial and temporal association among pixel arrays. For any pixel array $i$, the conditional distribution of $Y(x_{i,t})$, given $\{y(x_{j,t}); j \neq i \}$ on all pixel arrays $j$ which belongs to a pre-defined neighborhood $\Omega_i$ of $i$ (i.e., $j \in \Omega_i$ and $j \neq i$), is a Gaussian given as follows:
\begin{eqnarray}
     Y(x_{i,t}) | \{Y(x_{j,t}); j \neq i, j \in \Omega_i\} \sim N\left( \rho_t \sum_{j\neq i}\frac{w_{i,j}(t)}{w_{i+}(t)}y(x_{j,t}), \sigma_{i,t}^2 \right)
   \label{eq:STCAR_conditional}
\end{eqnarray}
where
\begin{eqnarray}
 w_{i,j}(t)=
    \begin{cases}
      \phi( x_{i,t}, x_{j,t}), & \text{if}\ j \in \Omega_i, i \neq j \\
      0, & \text{otherwise}
    \end{cases}
    \label{eq:w}
\end{eqnarray}
and $w_{i+}(t) = \sum_{j}^{n} w_{i,j}(t)$, and $\phi$ is a non-increasing function of the distance between $x_{i,t}$ and $x_{j,t}$. In (\ref{eq:STCAR_conditional}), the parameter $\sigma_{i,t}$ describes the variability of the data at location $x_{i,t}$, and the parameter $\rho_t$ represents the strength of spatial association at time $t$.

Note that, the distance between any two pixel arrays gradually changes over time. If the neighborhood of a pixel array $i$ is defined in a conventional way as a set of pixel arrays within a fixed distance of array $i$, then, the number of pixel arrays in a neighborhood varies over time as the pixel arrays travel in space. This is mathematically inconvenient for an autoregressive model like (\ref{eq:STCAR_conditional}). Fortunately, within a short period of time (say, 10 to 15 minutes), it is reasonable to assume that neighboring pixels tend to move in a similar direction with a similar speed, provided that the wind field is smooth and slowly varying in space. In fact, in the numerical example presented in Section \ref{sec:case}, the order of $q$ of the AR process (\ref{eq:AR}) is typically 2 to 3, which corresponds to 10 to 15 minutes. Based on the above considerations, we hence define the neighborhood of a pixel array $i$ using the initial locations of pixel arrays at time $t=1$ as follows:
 \begin{eqnarray}
 \Omega_i = \{ j \in \mathbb{S}: ||x_{i,1}-x_{j,1}||<d \}
\end{eqnarray}
for a constant distance $d>0$.

Let $\textbf{W}_t = \{ w_{i,j}(t) \}_{i,j=1}^{n}$, $\textbf{W}_{D,t}=\textrm{diag}(w_{1+}(t),w_{2+}(t),...,w_{n+}(t))$ and $\sigma_{i,t}=\sigma_t \cdot w_{i+}^{-1}(t)$, $\textbf{Y}_{t}$ is modeled as a temporal sequence of Conditional Auto-Regressive (CAR) models with expected value zero, i.e.,
\begin{eqnarray}
     \textbf{Y}_{t} \sim N\left(\textbf{0}, \sigma_t^2 (\textbf{W}_{D,t} - \rho_t \textbf{W}_t)^{-1}  \right).
   \label{eq:STCAR1}
\end{eqnarray}

Then, the vector of random fields, $\textbf{Y}_{t}$, is an STCAR model of order $q$ if for every $t$,
\begin{eqnarray}
     \textbf{B}_{t}\textbf{Y}_{t} = \sum_{j}^{q} r_j \textbf{B}_{t-j}  \textbf{Y}_{t-j} + \epsilon_t,
   \label{eq:STCAR2}
\end{eqnarray}
where
\begin{eqnarray}
     \epsilon_t \sim N\left(\textbf{0}, \sigma_t^2 \textbf{W}_{D,t}^{-1} (\textbf{I} - \rho_t \textbf{W}_{D,t}^{-1}\textbf{W}_t)^{T}  \right),
   \label{eq:STCAR3}
\end{eqnarray}
\begin{eqnarray}
     \textbf{B}_{t} = \textbf{I}-\rho_t \textbf{W}_{D,t}^{-1} \textbf{W}_t.
   \label{eq:STCAR4}
\end{eqnarray}
%\end{STCAR}

To ensure that the covariance matrix $\bm{\Sigma}_t$ is positive definite, $\rho_t$ needs to be between the interval $(1/\lambda^{\mathrm{max}}, 1/\lambda^{\mathrm{min}})$ with $\lambda^{\mathrm{min}}$ and $\lambda^{\mathrm{max}}$ respectively the maximum and minimum eigenvalues of the matrix $\bm{W}_{D,t}^{-1} \bm{W}_t$ (Haining 1990).

Substituting (\ref{eq:Gt}) and (\ref{eq:MeanFunction}) into (\ref{eq:AR}), we have
\begin{eqnarray}
  \bm{B}_{t}\textbf{G}_{t} = \sum_{j=1}^{j=q} r_j \bm{B}_{t-1} \textbf{F}_{t-j}\gamma_{t-j} + \sum_{j=1}^{j=q} r_j \bm{B}_{t-j} \textbf{Y}_{t-j} + \epsilon_t.
   \label{eq:AR2}
\end{eqnarray}

Hence, (\ref{eq:AR2}) suggests that the reflectivity growth process $\{\bm{G}_t\}$ is modeled by an STCAR with order $q$. The first term on the right hand side of (\ref{eq:AR2}) describes the mean process, while the second term on the right of (\ref{eq:AR2}) is an STCAR model defined by (\ref{eq:STCAR2}).

The autoregressive nature of the model (\ref{eq:AR2}) allows us to extrapolate the radar images for nowcasting purposes. At any time ${t}'$, the one-step-ahead reflectivity $Z_{t'+1}$ can be predicted by (\ref{eq:approximation}), i.e., $\textbf{Z}_{t'+1} = \textbf{Z}_{t'-1}+2\textbf{G}_{t'}$. Here, the reflectivity $\textbf{Z}_{t'-1}$ at $t'-1$ is observed, and $\textbf{G}_{t'}$ can be estimated from (\ref{eq:AR2}). In particular,
\begin{eqnarray}
  \textbf{G}_{t'}  \sim N \left( \sum_{j=1}^{q} {r}_j {\textbf{B}}_{t'}^{-1} {\textbf{B}}_{t'-j}\textbf{G}_{t'-j}, {\sigma}_{t'}^{2}(\textbf{W}_{D,t'}-{\rho}_{t'}\textbf{W}_{t'})^{-1} \right).
  \label{eq:prediction}
\end{eqnarray}
Similarly, the reflectivity fields at times $t'+2$, $t'+3$, ..., $t'+p$ are obtained iteratively.

\subsection{Parameter Estimation}\label{sec:estimation}
The proposed STCAR model contains a set of unknown parameters, including a $3J \times 1$ column vector, $\gamma_{t}=(\gamma_{1,t}^T,\gamma_{2,t}^T,...,\gamma_{J,t}^T)^T$, that determines the overall spatial trend of the reflectivity growth at time $t$, the spatial association $\rho_t$ at time $t$, the variability $\sigma_t$ at time $t$, and the parameters of the temporal association, $\textbf{r}=(r_{1},r_{2},...,r_{q})$.

Hence, a large number of unknown parameters is to be estimated for any time $t$. In the application example presented in the next section, for example, the number of kernels $J$ is chosen to be 30, which makes $\gamma_{t}$ a $90 \times 1$ column vector. In practice, a new radar image becomes available every 5 minutes. Considering the time consumed by the preliminary radar data processing and the construction of velocity field, it is of vital importance for the parameter estimation procedure to be numerically stable and efficient enough to incorporate the latest information into the analysis. This motivates us to adopt a two-step estimation approach. In the first step, we ignore the temporal association between radar images, and estimate the parameters $\gamma_{t}$, $\rho_t$, and $\sigma_t$ from individual radar images. In the second step, the temporal association $\textbf{r}=(r_{1},r_{2},...,r_{q})$ is estimated based on the results from the first step.

In particular, the parameters $\gamma_{t}$, $\rho_t$, and $\sigma_t$ are estimated using the Iteratively Re-Weighted Generalized Least Squares (IRWGLS). Note that, when the temporal association of radar images is ignored, the model (\ref{eq:AR2}) degenerates to a conventional CAR model without temporal correlation between a sequence of radar scans:
\begin{eqnarray}
     \textbf{B}_{t}\textbf{G}_{t} = \textbf{F}_{t}\gamma_{t} + \epsilon_t, \quad \epsilon_t \sim N\left(\textbf{0}, \sigma_{t}^2 \textbf{W}_{D,t}^{-1} (\textbf{I} - \rho_t \textbf{W}_{D,t}^{-1} \textbf{W}_t)^T \right).
   \label{eq:9}
\end{eqnarray}

The IRWGLS consists of the following steps:

\begin{description}
  \item[Step 0]:
  Set the initial $\hat{\bm{\Sigma}}_{t}$ to an identify matrix.
  \item[Step 1]:
  Estimate $\gamma_t$ using the Feasible General Least Squares (FGLS):
    \begin{equation*}
      \hat{\gamma}_t = (\bm{F}_t^{\top} \hat{\bm{\Sigma}}_{t}^{-1} \bm{F}_t)^{-1} \bm{F}_t^{\top} \hat{\bm{\Sigma}}_{t} \bm{G}_t.
    \end{equation*}
  \item[Step 2]:  Based on the residuals $ \bm{u}= \bm{G}_t - \bm{F}_t \hat{\gamma}_t$, estimate the spatial association $\rho_t$ and variability $\sigma_t$ using the Maximum Likelihood Estimation (MLE) described below, and obtain the estimate of the covariance matrix, $\hat{\bm{\Sigma}}_{t}$, from equation (\ref{eq:STCAR1}).
  \item[Step 3]: Iterate Steps 1 and 2 until the relative changes of $\hat{\gamma}_t$, $\hat{\rho}_t$ and $\hat{\sigma}_t$ are small. In the first iteration, since $\hat{\bm{\Sigma}}_{t}$ is an identify matrix, $\hat{\gamma}_t$ in Step 1 is the Ordinary Least Squares (OLS) estimator and is unbiased. In subsequent iterations, the finite-sample properties of the FGLS estimator, $\hat{\gamma}_t$, are usually unknown. Asymptotically, the FGLS estimator possesses the asymptotic properties of the Maximum Likelihood estimator, and is equivalent to the Generalized Least Squares (GLS) estimator under regularity conditions. In fact, it is possible to obtain estimate $\gamma_{t}$, $\rho_t$, and $\sigma_t$ all at once using MLE. However, this leads to a high-dimensional optimization problem which could be numerically inefficient in practice.
\end{description}

In step 2 above, given the observed reflectivity $\textbf{z}_{t-1}$ and $\textbf{z}_{t+1}$, the log-likelihood function is given as follows:
\begin{eqnarray}
    %l(\textbf{g}_{t}^{(i)};\beta,\gamma)=
    l(\textbf{y}_{t};\gamma_t,\sigma_t,\rho_t) =-\frac{n}{2}\log(2\pi\sigma_{t}^2)+\frac{1}{2}\log|\textbf{W}_{D,t}\textbf{B}_{t}| - \frac{\textbf{y}_{t}^T \textbf{W}_{D,t} \textbf{B}_{t}\textbf{y}_{t}}{2\sigma_{t}^2}.
   \label{eq:like1}
\end{eqnarray}
where
\begin{eqnarray}
    \textbf{y}_{t} = \textbf{g}_{t} - \textbf{F}\gamma_{t},
   \label{eq:like2}
\end{eqnarray}
and $\textbf{g}_{t} = (\textbf{z}(\textbf{x}_{t+1}) - \textbf{z}(\Gamma^{(-2)}\textbf{x}_{t+1})/2$.

Note that, the MLE of $\sigma_{t}^2$ is given by
\begin{eqnarray}
    \hat{\sigma}_{t}^2 = \frac{\textbf{y}_{t}^T \textbf{W}_{D,t} \textbf{B}_{t} \textbf{y}_{t}} {n}.
   \label{eq:like3}
\end{eqnarray}

Substituting (\ref{eq:like3}) into the log-likelihood function (\ref{eq:like1}), the MLE of $\gamma_t$ and $\rho_t$ can be numerically found by minimizing
\begin{eqnarray}
     \frac{n}{2} \log \left(  \frac{\textbf{y}_{t}^T \textbf{W}_{D,t} \textbf{B}_{t} \textbf{y}_{t}} {n} \right) - \frac{1}{2} \log|\textbf{W}_{D,t}\textbf{B}_{t}|.
   \label{eq:like4}
\end{eqnarray}
Note that, multiple initial values can be used to avoid the convergence to local minima, given the high dimension of the optimization problem.

After the parameters $\gamma_{t}$, $\rho_t$, and $\sigma_t$ have been estimated, it is possible to estimate the temporal association parameter $\textbf{r}=(r_{1},r_{2},...,r_{q})$. Note that, the model (\ref{eq:AR2}) degenerates to a conventional linear regression model (\ref{eq:lr}) without spatial correlation between the reflectivity on pixel arrays if $\rho_{t}=0$ for all $t$. Hence, $\textbf{r}$ can be estimated using the Weighted Least Squares after substituting $\hat{\gamma}_t$, $\hat{\sigma}_t$, $\hat{\textbf{W}}_{D,t}$ and $\textbf{y}_{t}$ into (\ref{eq:lr}):
\begin{eqnarray}
   \textbf{G}_{t} = \sum_{j=1}^{j=q} r_j \textbf{F}_{t-j} \hat{\gamma}_{t-j} + \sum_{j=1}^{j=q} r_j \textbf{Y}_{t-j} + \epsilon_t,
   \label{eq:lr}
\end{eqnarray}
where $\epsilon_t \sim N (\textbf{0}, \hat{\sigma}_{t}^2 \hat{\textbf{W}}_{D,t}^{-1} )$. Note that, $\bm{B}_{t}$ becomes an identity matrix if $\rho_{t}=0$.

%It follows immediately from the STCAR model (\ref{eq:5}) that the conditional distribution of $\textbf{G}_{t}$ given $(\textbf{G}_{t-1},\textbf{G}_{t-2},...,\textbf{G}_{t-q})$ can be written as
%\begin{eqnarray}
%     \textbf{G}_{t}|(\textbf{G}_{t-1},\textbf{G}_{t-2},...,\textbf{G}_{t-q}) \sim N\left( \sum_{i}^{q} r_i \textbf{B}_{t}^{-1} \textbf{B}_{t-i}\textbf{g}_{t-i},  \sigma_t^2 (\textbf{W}_{D,t} - \rho_t \textbf{W}_t)^{-1}   \right)
%   \label{eq:14}
%\end{eqnarray}
%
%Hence, the reflectivity growth in the next $p$ time periods, $(\hat{\textbf{G}}_{t},\hat{\textbf{G}}_{t+1},...,\hat{\textbf{G}}_{t+p-1})$, can be predicted iteratively using equation (\ref{eq:14}), which allows us to achieve the ultimate goal: predict the reflectivity on all pixel arrays, $(\hat{\textbf{Z}}_{t+1},\hat{\textbf{Z}}_{t+2},...,\hat{\textbf{Z}}_{t+p})$, in the next $p$ time periods.

\section{Application Example} \label{sec:case}
We revisit the motivating example presented in Section \ref{sec:Data}, and apply the proposed modeling approach to the radar images, shown in Figure \ref{fig:dBZcase}, on the early morning of 25 June 2010.

\subsection{Obtain the Velocity Field}
\begin{figure}
\begin{center}
\includegraphics[width=0.9\textwidth]{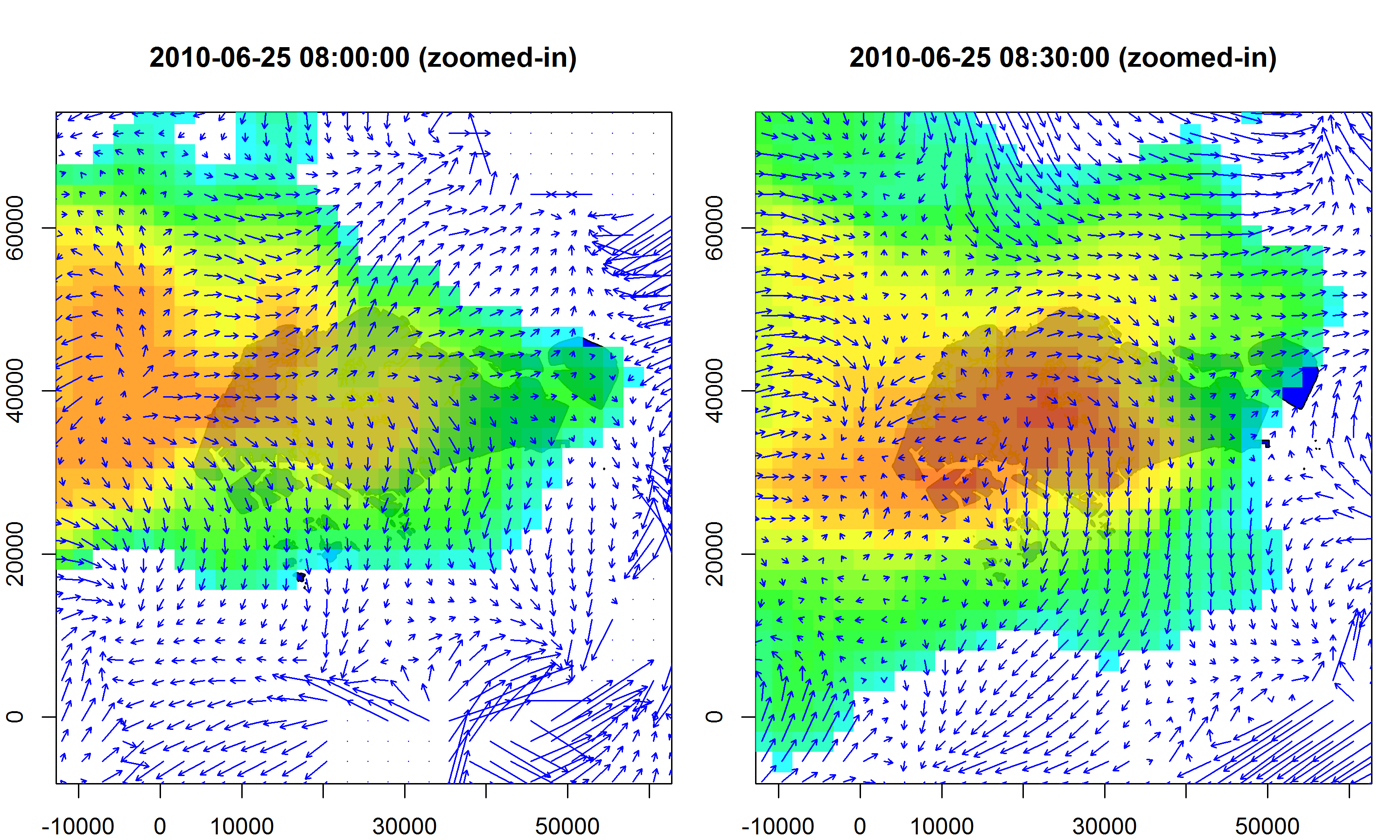}
\caption{Constructed velocity fields at 8:05AM and 8:30AM over Singapore. The arrows indicate the wind direction and the length of the arrow is proportional to the wind speed.} \label{fig:windfield}
\end{center}
\end{figure}

The first step is to construct the velocity field, making it possible to track pixel arrays as well as the reflectivity change on those arrays. For illustrative purposes, Figure \ref{fig:windfield} shows the constructed velocity fields at 8:00AM and 8:30AM, which are respectively the start and end times of the sequence of radar images shown in Figure \ref{fig:dBZcase}. The velocity fields are obtained using two consecutive radar scans. In order to visualize the velocity vectors clearly, we only show the constructed velocity fields over Singapore, i.e., the central part of the spatial domain. The moving directions of pixel arrays are indicated by arrows with their length proportional to the moving speed. Note that, since the velocity vectors are obtained based on the Pearson's correlation coefficient of the reflectivity values on pixel arrays from two consecutive scans, it is only possible to obtain the velocity vectors for areas where the weather system is located. In practice, of interest is always the motion of the weather system.

\begin{figure}
\begin{center}
\includegraphics[width=0.9\textwidth]{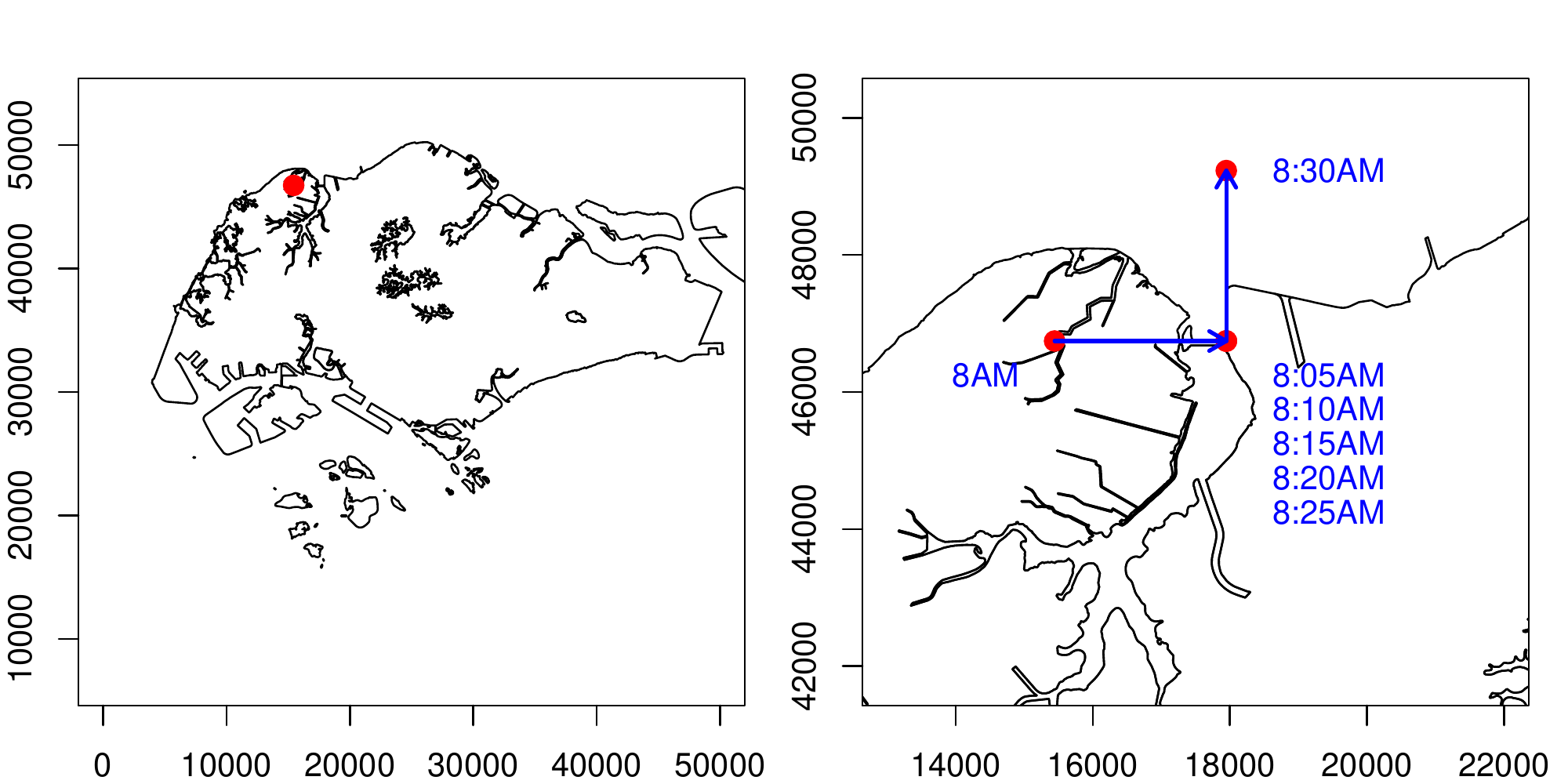}
\caption{Illustration of the tracking of pixel arrays. The left panel shows the initial location of a pixel array, and the right panel shows the tracked locations of this pixel array at different time.} \label{fig:tracking}
\end{center}
\end{figure}

\begin{figure}
\begin{center}
\includegraphics[width=0.8\textwidth]{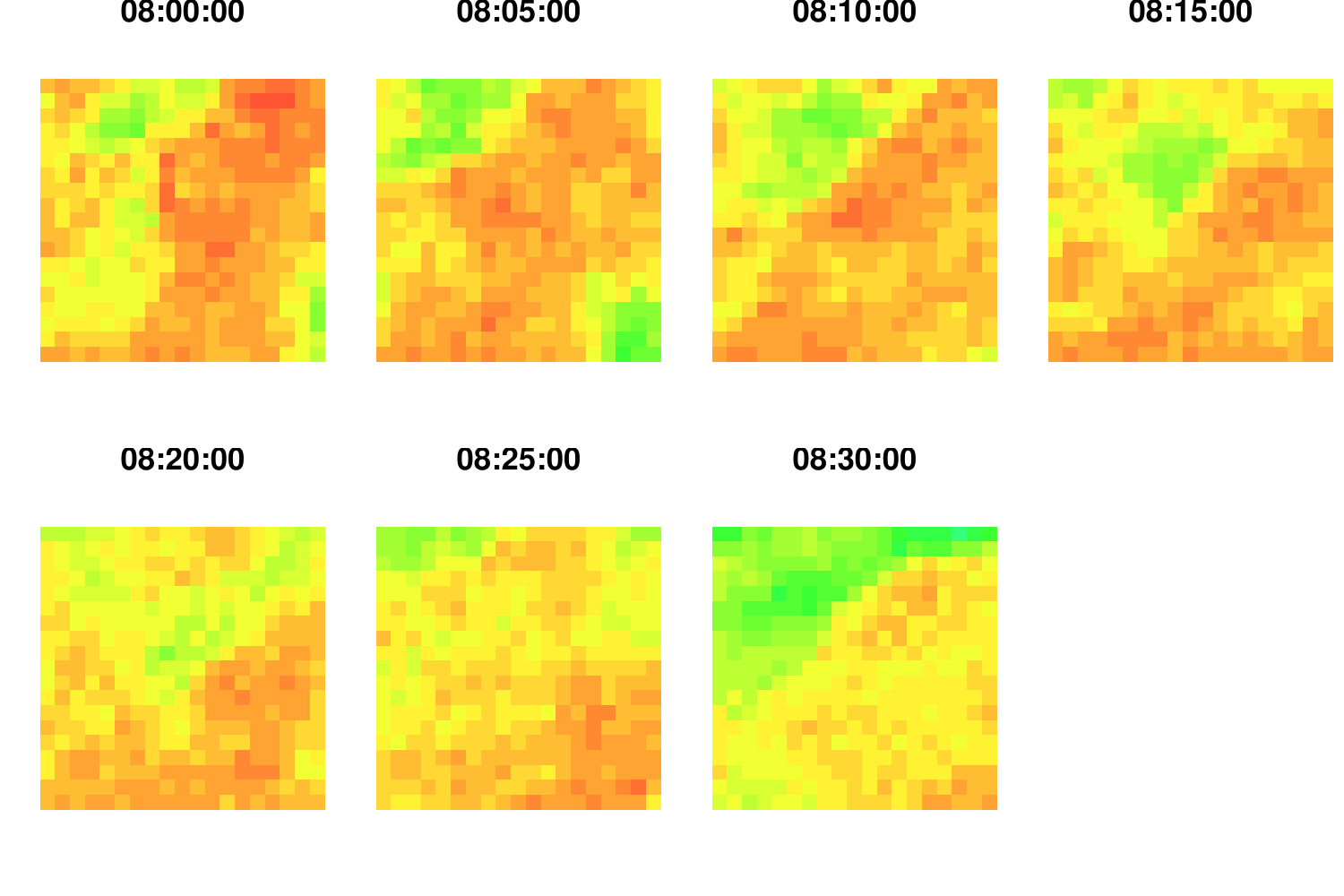}
\caption{Reflectivity values on this tracked pixel array at different times} \label{fig:trackingarray}
\end{center}
\end{figure}

To further illustrate the pixel tracking method, the left panel of Figure \ref{fig:tracking} shows the initial location of a pixel array with id number 2773 at 8:00AM. Recall that, there are 8649 pixel arrays in total. The right panel of Figure \ref{fig:tracking} shows the tracked locations of this pixel array at different times obtained from the cross-correlation-based method. This pixel array heads to the east at the beginning, and then turns to the north direction at 8:30AM. Figure \ref{fig:trackingarray} also shows the reflectivity on this tracked pixel array at different times. Recall that, each pixel array contains 361 pixels. It is seen that the seven reflectivity images share similar patterns, and are hence connected.

\subsection{Model the Reflectivity Field}
\begin{figure}
\begin{center}
\includegraphics[width=0.95\textwidth]{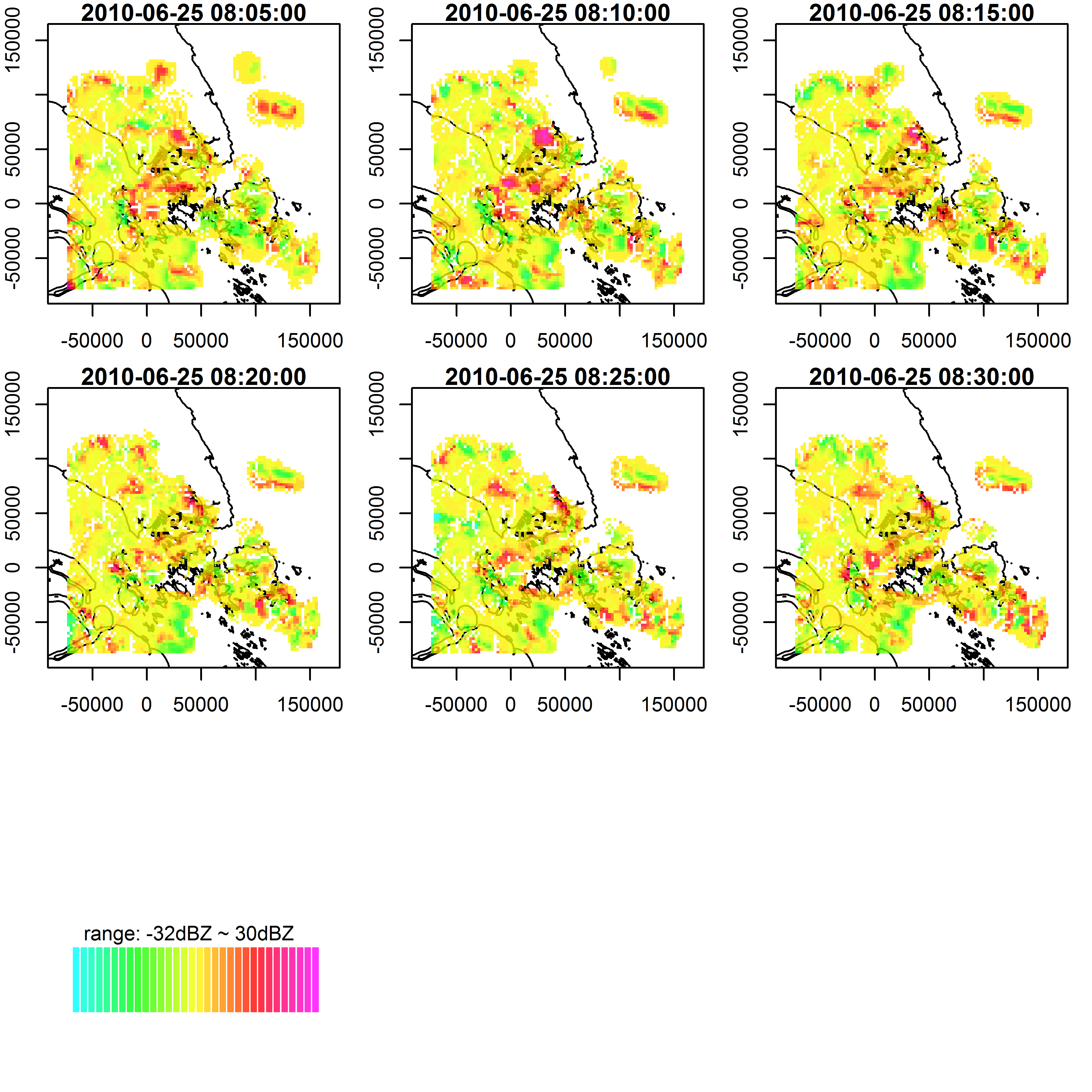}
\caption{Reflectivity growth over the spatial domain from 8:05AM and 8:30AM} \label{fig:growth}
\end{center}
\end{figure}

Once the velocity field has been established, the reflectivity growth, $\textbf{G}_t$, is obtained using (\ref{eq:growth}) and shown in Figure \ref{fig:growth}. As one might expect, high reflectivity growth is found at the central area of the storm where the weather system is highly dynamic, while the reflectivity approximately remains unchanged over the 5-min sampling interval for most pixel arrays.

The model parameters $\rho$, $\sigma$ and $r$ are estimated using the method described in Section \ref{sec:estimation}. In particular,
%the function $\phi$ is chosen to be $\exp(-0.1d_{i,j})$ where $d_{i,j}$ is the distance between pixel arrays $x_{i,t}$ and $x_{j,t}$ in kilometer.
we place a number of $J=30$ Gaussian kernels over the reflectivity field with the centers chosen by the method of K-means clustering (Hartigan and Wong 1979). For each Gaussian kernel, the covariance matrix is chosen as a diagonal matrix with standard deviation $10\mathrm{km}$. Table \ref{table:rhosigma} shows the estimated $\rho$ and $\sigma$ from 8:05AM to 8:30AM.

\begin{table}[h!]
   \centering
   \caption{Estimated values of $\rho$ and $\sigma$ from 8:05AM to 8:30AM} \label{table:rhosigma}
\begin{tabular}{ c || c | c | c | c | c | c  }
   \hline
    & 8:05AM & 8:10AM & 8:15AM & 8:20AM & 8:25AM & 8:30AM \\
   \hline
   $\rho$ & -11.63 & -19.22 & -18.96 & -19.12 & -19.19 & -19.32 \\
   $\sigma$ & 18.67 & 16.20 & 16.34 & 15.84 & 17.44 & 17.08 \\
   \hline
\end{tabular}
\end{table}

\begin{figure}
\begin{center}
\includegraphics[width=1\textwidth]{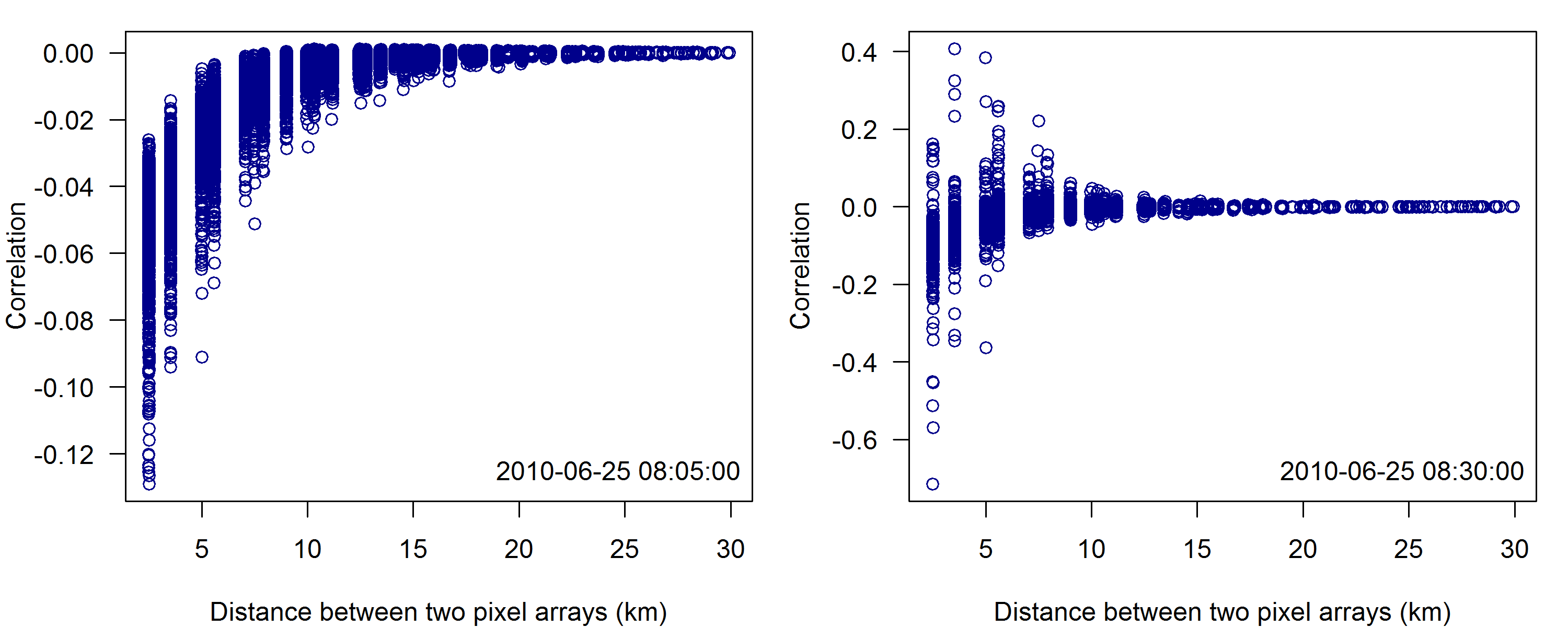}
\caption{Smoothed correlation between all pairs of pixel arrays} \label{fig:correlation}
\end{center}
\end{figure}

As seen from Table \ref{table:rhosigma}, the values of $\sigma$, which characterizes the overall variability, does not change dramatically over the 25-min interval from 8:05AM to 8:30AM. The spatial association $\rho$, however, jumps from $-11.63$ at 8:05AM to $-19.22$ at 8:10AM and remains around $-19$ after that. Substituting the estimated values from Table \ref{table:rhosigma} into (\ref{eq:STCAR1}), we obtain the estimated covariance matrix $\hat{\bm{\Sigma}}_t$, which allows us to further investigate the correlation structure of the reflectivity growth among neighboring pixel arrays. Figure \ref{fig:correlation} plots the computed correlation between all pairs of pixel arrays against their distances in kilometer. In particular, the left panel shows the correlation at 8:05AM where $\rho$ is close to $-11$, while the right panel corresponds to 8:30AM where $\rho$ is close to $-19$. Interestingly, the two plots suggest very different correlation structures. At 8:05AM, the neighboring pixel arrays are negatively correlated. For two pixel arrays which are close to each other, some pairs of pixel arrays exhibits stronger correlation (say, $-0.12$) than others (say, 0.02). In general, the correlation becomes weaker as the distance between two pixel arrays grows and can be ignored when two pixel arrays are 30km apart. At 8:30AM, the correlation structure among pixel arrays seems to be more complicated. Not only the strength of correlation becomes much stronger $(-0.6 \sim 0.4)$, but also the neighboring pixel arrays could either exhibits positive or negative correlation. Such a puzzling result concerning the correlations implied by the CAR model has been studied by Wall (2004) and Assuncao and Krainski (2009). The authors found that, when $\rho$ increases from zero to the upper bound of its parameter space, the correlation is positive and monotone increasing as $\rho$. When $\rho$ decreases from zero, the correlation is negative at first and also decreases as $\rho$. But, when $\rho$ is further approaching to the lower bound of its parameter space, the correlation between some neighbors could either approaches $-1$ or starts growing to the positive side. In our context, the negative correlation among two neighboring pixel arrays might be explained by the mass conservation as the water contained in the cloud leaving one region and joining its neighboring regions. The positive correlation is often expected as the reflectivity in a particular region is growing or decaying at the same time.

%Given the dynamics of the storm system, the temporal association $r$ also needs to be updated dynamically in the sense the estimated values are always obtained from the most recent radar scans. Table \ref{table:r} shows the estimated $r$ based on an STCAR of order 3 (i.e., $q=3$).
%
%\begin{table}[h!]
%   \centering
%   \caption{Test} \label{table:r}
%\begin{tabular}{| c || c | c | c | }
%   \hline
%    & $r_1$ & $r_2$ & $r_3$  \\
%   \hline
%   8:20AM & 0.17 & -0.34 & 0.04  \\
%   8:25AM & 0.31 & -0.45 & 0.03   \\
%   8:30AM & 0.26 & -0.38 & 0.03   \\
%   \hline
%\end{tabular}
%\end{table}

\begin{figure}
\begin{center}
\includegraphics[width=0.9\textwidth]{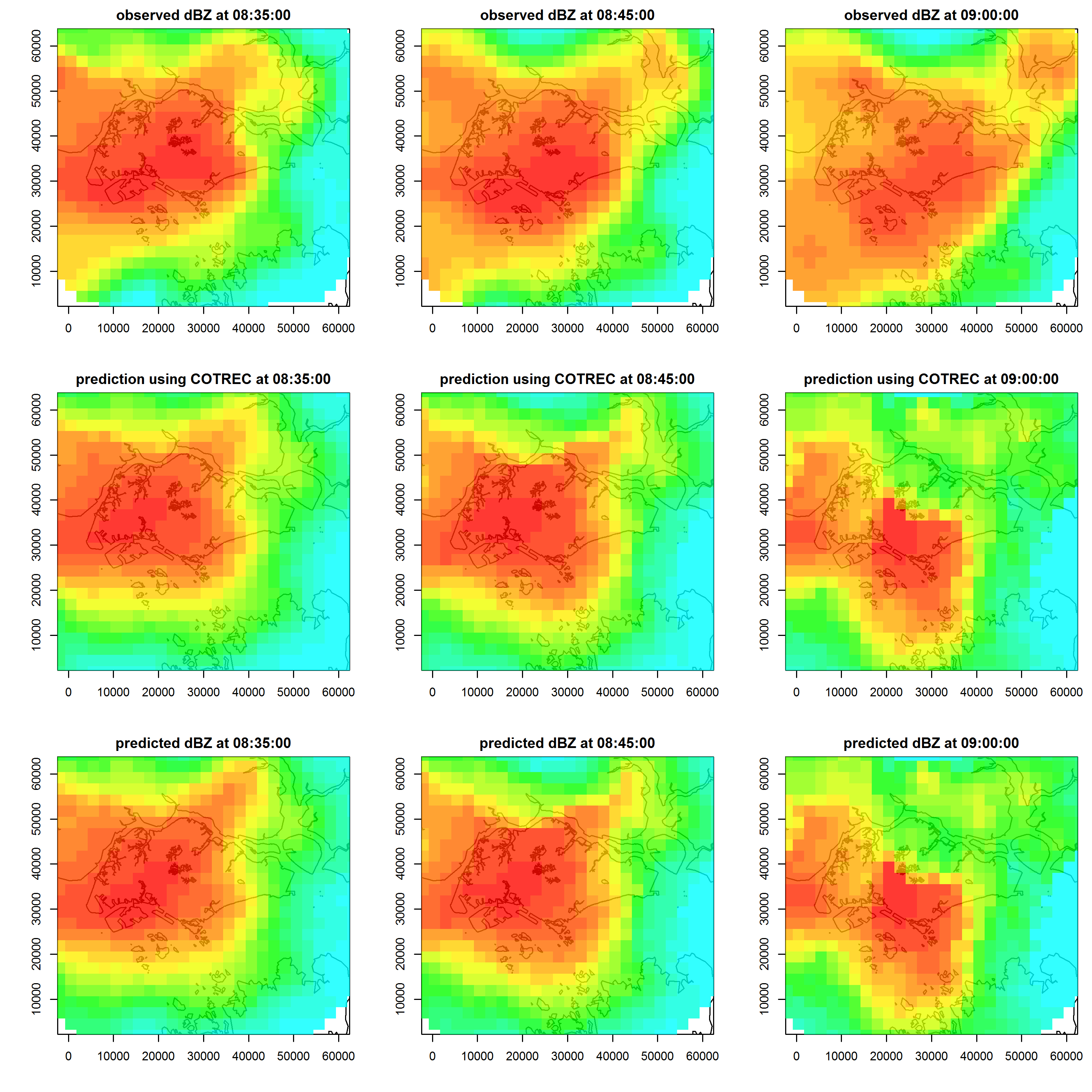}
\caption{Observed and predicted reflectivity fields at 8:35AM, 8:45AM and 9:AM. The first row shows the actually observed reflectivity, the second rows shows the predicted reflectivity using the existing COTREC system, and the third row shows the predicted reflectivity using the proposed approach} \label{fig:predict}
\end{center}
\end{figure}

Once $\hat{\gamma}_t$, $\hat{\sigma}_t$ and $\hat{\textbf{W}}_{D,t}$ has been estimated, $\bm{r}$ is obtained from (\ref{eq:lr}) using the Weighted Least Squares. If the order of the STCAR model is chosen to be $q=2,3,4$, we respectively obtain $\hat{\bm{r}}=(0.911,-0.317)$, $\hat{\bm{r}}=(0.996, -0.495, 0.21)$ and $\hat{\bm{r}}=(1.082, -0.513, 0.347, -0.206)$. The autoregressive nature of the model (\ref{eq:AR2}) allows us to extrapolate the radar images for nowcasting purposes. The reflectivity field at 8:35AM (i.e., 5-minute-ahead nowcasting) is predicted using (\ref{eq:prediction}) and is shown in Figure \ref{fig:predict}. Similarly, the reflectivity fields at 8:15AM (i.e., 15-minute-ahead) and 9:00AM (i.e., 30-minute-ahead) are also obtained iteratively and shown in Figure \ref{fig:predict}. In this figure, the first row shows the actually observed reflectivity at 8:35AM, 8:45AM and 9:00AM, the second rows shows the predicted reflectivity using the existing COTREC system (Li and Lai 2004), and the third row shows the predicted reflectivity using the proposed approach.

%\subsection{Comparison Studies}

The COTREC method is a well-known QPF approach which has been widely implemented in the meteorological society (Li and Lai 2004, RMI 2008). While both the COTREC and the proposed approaches rely on the same pattern-based method to construct the wind field, the former (which is not a statistical model) does not model the growth and decay of the reflectivity as the proposed method does. Hence, the COTREC method serves as an ideal candidate for comparison purposes. It is seen from Figure \ref{fig:predict} that, the reflectivity fields generated by both methods reasonably match with the actually observed reflectivity fields for 5-min- and 15-min-ahead nowcasting. However, the 30-min-ahead nowcasting at 9:00AM becomes less accurate. At 9:00AM, the actual reflectivity field heads to the east, while the predicted high reflectivity region moves to the south. Since the radar-based nowcasting method is based on the spatio-temporal extrapolation (autoregresion in our case) of the radar reflectivity, it is extremely difficult to predict the change of the wind direction for highly dynamic tropical storms. Note that, the constructed velocity vectors associated with the high reflectivity region at the center of the spatial domain (as shown by Figure \ref{fig:windfield}) are pointing to the south at 8:30AM. In general, the predicability of the same radar-based nowcasting method varies with geo-locations (Radhakrishna \emph{et al.} 2012). In tropical areas where storm systems are highly dynamic and embedded with numerous small-scale convective cells which change rapidly within tens of minutes, the radar-based nowcasting methods usually provide reasonable quantitative prediction within a very short period of time. For longer period prediction, it is a common practice in the meteorological society to incorporate the downscaled output generated by computationally-intensive numerical weather prediction models, such as the Weather Research and Forecast (WRF) model, through data assimilation (Bowler \emph{et al.} 2006).

\begin{figure}
\begin{center}
\includegraphics[width=1\textwidth]{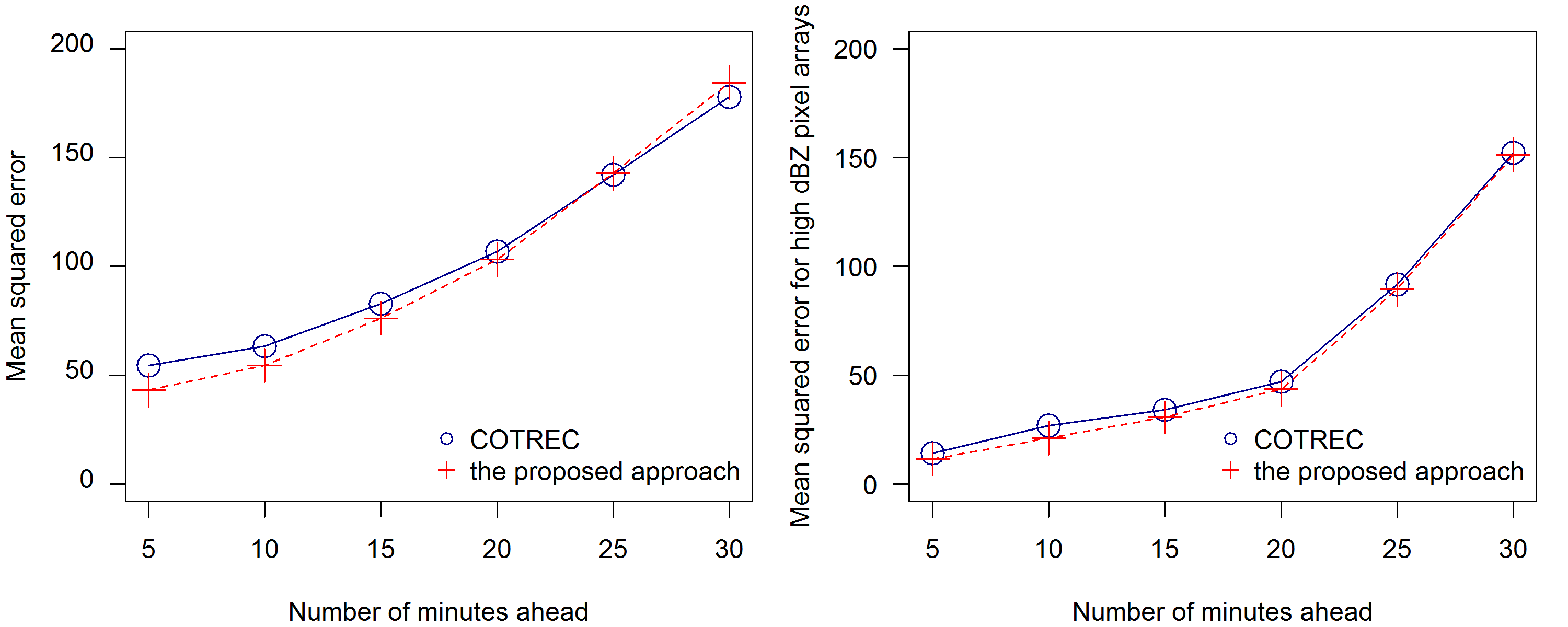}
\caption{The comparison of MSE between the existing COTREC method and the proposed method. The left panel shows the MSE computed from all pixel arrays, and the right panel shows the MSE computed only from those pixel arrays of interest with reflectivity greater than 35dBZ} \label{fig:comparison}
\end{center}
\end{figure}

To further compare the performance between the COTREC and the proposed method, Figure \ref{fig:comparison} shows the accumulative mean-squared-error of the predicted reflectivity at pixel arrays. Note that, we focus on the accumulative MSE rather than the MSE at each time point because precipitation nowcasting in practice is often concerned with the prediction of the accumulative amount up to a certain time point. The left panel shows the comparison for all pixel arrays. In addition, since only the pixel arrays with high reflectivity are of interest, the right panel shows the comparison for pixel arrays with reflectivity greater than 35dBZ. Empirically, 35dBZ corresponds to $5.6\mathrm{mm}/\mathrm{hr}$ or $0.22\mathrm{in}/\mathrm{hr}$ moderate rains. We see that, it is worth considering the growth and decay of the reflectivity for short-term nowcasting. As seen from the left panel of Figure \ref{fig:comparison}, the proposed method outperforms if the nowcasting horizon is not greater than 25 minutes. If we only focus on pixel arrays with reflectivity greater than 35dBZ, the proposed method outperforms up to 30-min-ahead nowcasting. As we have already discussed, the reflectivity growth is extrapolated through a space-time autoregression model, the performance of such a model will deteriorate as the prediction horizon increases for extremely dynamic tropical storm. Hence, our findings suggest that the gain of incorporating the growth and delay into the model becomes zero or even negative when the prediction horizon is beyond what the model can offer. The cut-off threshold varies with geo-locations, seasons and types of precipitation, and can be found in practice through numerical experiments on historical data.

\section{Conclusions} \label{sec:conclusion}

A spatio-temporal approach has been proposed for modeling a sequence of radar images. Given the velocity vector field, i.e, the motion field, of the weather system, the spatio-temporal process of the radar reflectivity was modeled by a space-time conditional autoregressive model. Compared to existing nowcasting methods available to the meteorological society, the proposed statistical approach is able to capture the stochastic growth and decay of the strength of reflectivity. A numerical example, based on a real dataset, was provided to demonstrate the application of the method. It has been shown that the proposed model provides more accurate predictions than existing operational systems. Note that, the paper is only concerned with the modeling of two-dimensional radar reflectivity fields at a fixed altitude (1km above the sea level), while the vertical motion of the weather system has not been considered. In fact, a weather system is a complex and dynamic three-dimensional object, and modern radar systems are equipped with the capability to return the reflectivity at multiple altitude layers simultaneously. Hence, one important future work is to explore the modeling of three-dimensional radar reflectivity fields by extending the proposed method.

%\section*{Acknowledgment}
%The authors would like to thank the National Environment Agency of Singapore for providing the environmental monitoring measurements used in this paper.

\appendix

\section{Appendix}  \label{sec:appendix}
The pattern-based method for constructing the velocity vector field as well as the translation operators is described. Let a two-dimension discrete random function $Z_t(k_1,k_2;s_i)$ represent the reflectivity at time $t$ on the $(k_1,k_2)\text{th}$ pixel within the pixel array centered at $s_i$. Here, $k1,k2=1,2,...,19$ as each pixel arrays consists of $19 \times 19$ pixels. The velocity vector at location $s_i$ is given by $s_{j^*}-s_i$ where $s_{j^*}$ is found by maximizing the Pearson's correlation coefficient, $r$, between $Y_t(\cdot,\cdot;s_i)$ and $Y_{t+1}(\cdot,\cdot;s_j)$, i.e.,
\begin{eqnarray}
\begin{split}
   j^* = \mathrm{argmax}_j = r \left( Y_t(\cdot,\cdot;s_i),Y_{t+1}(\cdot,\cdot;s_j)  \right)
   \label{eq:trec}
\end{split}
\end{eqnarray}

In other words, the Pearson's correlation coefficient between the reflectivity patterns within pixel arrays is computed for all possible pairs of pixel arrays in two successive radar scans respectively taken at times $t$ and $t+1$. Then, the pair with the highest correlation is connected forming the velocity vector. One key assumption behind this method is that the shape of the reflectivity image within a pixel array does not change dramatically, which is valid when time interval is short, say, 5min as in this paper.

Let $\tilde{v}_t(s_i)=(\tilde{v}_t^{(1)}(s_i), \tilde{v}_t^{(2)}(s_i))^T$ be the velocity vector at location $s_i$ obtained from equation (\ref{eq:trec}), where $\tilde{v}_t^{(1)}$ and $\tilde{v}_t^{(2)}$ are the horizontal and vertical components of the velocity vector. To improve the consistency of the constructed velocity field, the smoothed velocity vectors are obtained by minimizing the sum of squared error
\begin{eqnarray}
\begin{split}
   J_1 = \int_{S}\left\{ (\tilde{v}_t^{(1)}-v_t^{(1)})^2+(\tilde{v}_t^{(2)}-v_t^{(2)})^2 \right\}dxdy
   \label{eq:7}
\end{split}
\end{eqnarray}
and subject to the Boussineq mass continuity equation
\begin{eqnarray}
\begin{split}
   \frac{\partial{v_t^{(1)}}}{\partial{x}} + \frac{\partial{v_t^{(2)}}}{\partial{y}}  = 0
   \label{eq:8}
\end{split}
\end{eqnarray}

It is known that the constrained minimization problem above is equivalent to the following unconstrained problem (Bertsekas 1982)
\begin{eqnarray}
\begin{split}
   J_2 = \int_{S}\left\{ (\tilde{v}_t^{(1)}-v_t^{(1)})^2+(\tilde{v}_t^{(2)}-v_t^{(2)})^2 + \lambda \left(\frac{\partial{v_t^{(1)}}}{\partial{x}} + \frac{\partial{v_t^{(2)}}}{\partial{y}} \right)\right\}dxdy
   \label{eq:9}
\end{split}
\end{eqnarray}
which can be efficiently solved using a variational analysis. Here, $\lambda$ is the Lagrangian multiplier.

Once the smoothed velocity vector $v_t$ has been found, we formally define the forward and inverse translation operators as follows:
\begin{eqnarray}
\begin{split}
   \Gamma\textbf{x}_t = \begin{pmatrix}x_{1,t} \\ x_{2,t} \\ ... \\ x_{n,t} \end{pmatrix} + \begin{pmatrix} v_t(x_{1,t}) \\ v_t(x_{2,t}) \\ ... \\ v_t(x_{n,t}) \end{pmatrix}= \textbf{x}_{t+1}
   \label{eq:10}
\end{split}
\end{eqnarray}
and
\begin{eqnarray}
\begin{split}
   \Gamma^{-1}\textbf{x}_{t+1} = \begin{pmatrix}x_{1,t+1} \\ x_{2,t+1} \\ ... \\ x_{n,t+1} \end{pmatrix} - \begin{pmatrix} v_t(x_{1,t}) \\ v_t(x_{2,t}) \\ ... \\ v_t(x_{n,t}) \end{pmatrix}= \textbf{x}_{t}.
   \label{eq:10}
\end{split}
\end{eqnarray}

%\bibliographystyle{asa}
%\bibliography{AQref}

\clearpage
\large{\textbf{References}}

National Envrionmental Agency Singapore (NEA) (2014), ``Weather Statistics'', available at http://app2.nea.gov.sg/weather-climate/climate-information/weather-statistics.

Royal Meteorological Institute of Belgium (RMI) (2008), ``Quantitative Precipitation Forecasts
based on radar observations: principles, algorithms and operational systems''.

Assuncao, R. and Krainski, E. (2009), ``Neighborhood Dependence in Bayesian Spatial Models'',
Biometrical Journal, 51, 851-869.

Banerjee, S., Carlin, B. P., and Gelfand, A. E. (2004), ``Hierarchical Modeling and Analysis for Spatial Data'', New York: Chapman and Hall/CRC.

Bertsekas, D. P. (1982), Constrained Optimization and Lagrange Multiplier Methods, Boston: Academic
Press.

Besag, J. and Kooperberg, C. (1995), ``On Conditional and Intrinsic Autoregression'', Biometrika,
82, 733-746.

Bowler, N. E., Pierce, C. E., and Seed, A. W. (2006), ``A Probabilistic Precipitation Forecasting
Scheme which Merges an Extrapolation Nowcast with Downscaled NWP'', Quarterly Journal of the Royal Meteorological Society, 132, 2127-2155.

Bowler, N. E. H., Pierce, C. E., and Seed, A. (2004),``Development of a precipitation nowcasting
algorithm based upon optical flow techniques'', Journal of Hydrology, 288, 74-91.

Brown, P. E., Diggle, P. J., Lord, M. E., and Young, P. C. (2001), ``Space-Time Calibration of
Radar Rainfall Data'', Journal of the Royal Statistical Society, Ser. C, 50, 221-241.

Browning, K. A. Collier, C. G. Lark, P. R. Menmuir, P. Monk, G. A. and Owens, R. G. (1982), ``On the
Forecasting of Frontal Rain Using a Weather Radar Network'', Monthly Weather Review, 110, 534-552.

Carlin, B. P. and Banerjee, S. (2003), Hierarchicla Multivariate CAR Models for Spatio-Temporally
Correlated Survival Data (with discussion), Oxford: Oxford University Press.

Cressie, N. (1993), Statistics for Spatial Data, New York: Wiley.

Fuentes, M., Reich, B., and Lee, G. (2008), ``Spatial-Temporal Mesoscale Modeling of Rainfall
Intensity Using Gage and Radar Data'', The Annals of Applied Statistics, 2, 1148-1169.

Gelpke, V. and Kunsch, H. R. (2001), ``Estimation of Motion from Sequences of Images'', in Spatial
Statistics: Methodological Aspects and Applications, ed. Moore, M., Springer, pp. 141-167.

Germann, U. and Zawadzki, I. (2002), ``Scale-Dependence of the Predicability of Precipitation from
Continental Radar Images. Part I: Description of the Methodology'', Monthly Weather Review,
130, 2859-2873.

Haining, R. (1990), Spatial Data Analysis in the Social and Environmental Sciences, Cambradge: Cambradge University Press.

Hartigan, J. A. and Wong, M. A. (1979), ``A K-Means Clustering Algorithm'', Applied Statistics, 28, 100-108.

Horn, B. K. P. and Schunck, B. G. (1981), ``Determining optical flow'', Artificial Intelligence, 17, 185-203.

Jin, X., Carlin, B. P., and Banerjee, S. (2005), ``Generalized Hierarchical Multivariate CAR Models
for Areal Data'', Biometrics, 61, 950-961.

Leese, J. A., Novak, C. S., and Clark, B. B. (1971), ``An automated technique for obtaining cloud
motion from geosynchronous saellite data using cross correlation'', Journal of Applied Meteorology, 10, 118-132.

Li, L., Schmid, W., and Joss, J. (1995), ``Nowcasting of motion and growth of precipitation with
radar over a complex orography'', Journal of Applied Meteorology, 34, 1286-1300.

Li, P. and Lai, S. T. (2004), ``Short-range quantitative precipitation forecasting in Hong Kong'', Journal of Hydrology, 288, 189-209.

Mariella, L. and Tarantino, M. (2010), ``Spatial Temporal Conditional Auto-Regressive Model: A New Autoregressive Matrix'', Austrian Journal of Statistics, 39, 223-244.

Marshall, J. S. and Palmer, W. M. (1948), ``The distribution of raindrops with size'', Journal of Meteorology, 5, 165-166.

Radhakrishna, B. Zawadzki, I. and Fabry, F. (2012), ``Predictability of Precipitation from Continental
Rada Images. Part V: Growth and Decay'', Journal of the Atmospheric Sciences, 69, 3336-3349.

Rinehart, R. E. and Garvey, E. T. (1978), ``Three-dimensional storm motion detection by conventional
weather radar'', Nature, 273, 287-289.

Sigrist, F., Kunsch, H., and Stahel, W. (2012), ``A dynamic nonstationary spatio-temporal model
for short term prediction of precipitation'', The Annals of Applied Statistics, 6, 1452-1477.

Staniforth, A. and Cote, J. (1991), ``Semi-Lagrangian Integration Schemes for Atmospheric Models: A Review'', Monthly Weather Review, 119, 2206-2223.

Stern, H. and Cressie, N. (2000), ``Posterior Predictive Model Checks for Disease Mapping Models'', Statistics in Medicine, 19, 2377-2397.

Stroud, J. R., Muller, P., and Sanso, B. (2001), ``Dynamic Models for Spatiotemporal Data'', Journal of the Royal Statistical Society, B, 63, 673-689.

Testik, F. and Gebremichael, M. (2013), Rainfall: State of the Science, American Geophysical
Union.

Wall, M. M. (2004), ``A Close Look at the Spatial Strucuture Implied by the CAR and SAR Models," Journal of Statistical Planning and Inference, 121, 311-324.

Wilson, J. W. Crook, N. A. Mueller, C. K. Sun, J. and Dixon, M. (1998), ``Nowcasting Thunderstorms: A
Status Report'', Bulletin of American Meteorological Society, 79, 2079-2099.

Wolfson, M. M. Forman, B. E. H. R. G. and Moore, M. P. (1999), ``The Growth and Decay Storm Tracker'', in The 8th Conference on Aviation, Range, and Aerospace Meteorology.

Xu, K., Wikle, C., and Fox, N. (2005), ``A kernel-based spatio-temporal dynamical model for nowcasting weather radar reflectivities'', Journal of the American Statistical Association, 100, 1133-1144.

\end{document}